\documentclass[sigconf,screen]{acmart}

\AtBeginDocument{%
  \providecommand\BibTeX{{%
    Bib\TeX}}}

\usepackage{amsmath,amsfonts}
\usepackage{algorithmic}
\usepackage{graphicx}
\usepackage{textcomp}
\usepackage{xcolor}
\usepackage{enumitem}
\usepackage{amsmath}
\usepackage{bbold}
\usepackage{comment}
\usepackage{graphicx}
\usepackage{hyperref}
\usepackage{verbatim}
\usepackage{cleveref}
\usepackage{listings}
\usepackage{booktabs}
\usepackage{subcaption}
\usepackage{framed}

\def\BibTeX{{\rm B\kern-.05em{\sc i\kern-.025em b}\kern-.08em
    T\kern-.1667em\lower.7ex\hbox{E}\kern-.125emX}}
\definecolor{light-gray}{gray}{0.80}

\graphicspath{{./figs/}}

\begin{document}

\copyrightyear{2026}
\acmYear{2026}
\setcopyright{cc}
\setcctype{by-nc-nd}
\acmConference[ICSE-SEIP '26]{2026 IEEE/ACM 48th International Conference on Software Engineering}{April 12--18, 2026}{Rio de Janeiro, Brazil}
\acmBooktitle{2026 IEEE/ACM 48th International Conference on Software Engineering (ICSE-SEIP '26), April 12--18, 2026, Rio de Janeiro, Brazil}
\acmPrice{}
\acmDOI{10.1145/3786583.3786901}
\acmISBN{979-8-4007-2426-8/2026/04}







\title[What's DAT? Three Case Studies of Measuring Software Development Productivity at Meta With Diff Authoring Time]{What's DAT? Three Case Studies of Measuring Software Development Productivity at Meta With Diff Authoring Time}


\author{Moritz Beller}
\orcid{0000-0003-4852-0526}
\email{mmb@meta.com}
\affiliation{%
  \institution{DevInfra, Meta Platforms, Inc.}
  \city{Menlo Park}
  \country{USA}
}

\author{Amanda Park}
\affiliation{%
  \institution{DevInfra, Meta Platforms, Inc.}
  \city{Menlo Park}
  \country{USA}
}

\author{Karim Nakad}
\affiliation{%
  \institution{DevInfra, Meta Platforms, Inc.}
  \city{Menlo Park}
  \country{USA}
}

\author{Akshay Patel}
\affiliation{%
  \institution{DevInfra, Meta Platforms, Inc.}
  \city{Menlo Park}
  \country{USA}
}

\author{Sarita Mohanty}
\author{Ford Garberson}
\affiliation{%
  \institution{DevInfra Data Science, Meta Platforms, Inc.}
  \city{Menlo Park}
  \country{USA}
}

\author{Ian G. Malone}
\author{Vaishali Garg}
\affiliation{%
  \institution{App Foundation Data Science, Meta Platforms, Inc.}
  \city{Menlo Park}
  \country{USA}
}

\author{Henri Verroken}
\affiliation{%
  \institution{DevInfra, Meta Platforms, Inc.}
  \city{London}
  \country{UK}
}

\author{Andrew Kennedy}
\affiliation{%
  \institution{DevInfra, Meta Platforms, Inc.}
  \city{London}
  \country{UK}
}

\author{Pavel Avgustinov}
\affiliation{%
  \institution{DevInfra, Meta Platforms, Inc.}
  \city{London}
  \country{UK}
}

\renewcommand{\shortauthors}{Beller et al.}

\begin{abstract}
This paper introduces Diff Authoring Time (DAT), a precise, yet conceptually simple approach to measuring software development productivity that enables rigorous experimentation. DAT is a time-based metric, which assesses how much active work time engineers take to develop self-contained changes. It uses a bespoke telemetry system integrated with version control, the Integrated Development Environments, and the Operating System. We validate DAT through observational studies, surveys, visualizations, and descriptive statistics.
At Meta, DAT elevates the internal tool development workflow to the scientifically grounded, experiment-driven development flow already present for external-facing products. As such, DAT enables rigorous experimentation on long-standing software engineering questions like ``do types make development more efficient?''
DAT has powered experiments and case studies on more than 20 projects at Meta. Here, we highlight
(1) a controlled experiment on introducing mock types, which showed that typed mocks in tests yield a 14\% DAT improvement,
(2) a case study on the development of automatic memoization in the React compiler, a 33\% improvement, and
(3) a framework to estimate the thousands of DAT hours saved annually through code sharing ($>50$\% improvement).
Overall, this paper contributes a novel, yet straightforward way to measure development velocity, DAT. It describes how the use of this metric has enabled rigorous experimentation on development productivity at industry scale, and paves the way toward a more scientific mindset in internal product development.
\end{abstract}

\maketitle

\section{Introduction}

For nearly two decades, Continuous Experimentation (CE) has been integral to external-facing product development in Big Tech~\cite{kohavi2009online,tang2010overlapping,karrer2021network}: Engineers frequently launch A/B experiments to evaluate the effects of new features. CE exposes a typically smaller cohort of users (A) to the proposed change, while withholding it from a control group (B), ensuring observed differences between A and B occur solely due to the proposed change and are free of interference. With a large user base, even small effect sizes can thus be detected.
Only when a change is shown to have a statistically significant positive effect on the outcome measure, will it generally roll out (to B, too).
In contrast to this rigorous, science-inspired development process for external products, is the development process of internal tools at the same companies often much less stringent: it mostly relies on a combination of basic usage numbers and qualitative feedback such as expert opinion, user surveys, anecdotes, and simply gut feel~\cite{forsgren2024devex}. This is puzzling given the tens of thousands of internal users and high value of engineering time, which constitute major operational expenses to the businesses.
To improve internal development, a scientific, metrics- and experimentation-driven culture shift for internal development is needed. However, assessing the ultimate topline of internal development, measuring productivity accurately, remains challenging~\cite{jaspan2019no,murphy2019predicts,beller2020mind}, as traditional metrics are often inaccurate and easily gameable.

In this paper, we propose a possible pragmatic solution for such a productivity metric and how it can change an entire organization's view on its development culture. We introduce Diff Authoring Time (DAT) as a key metric for assessing development velocity. DAT measures the active work time it takes engineers to develop a change, using a telemetry system integrated with version control, the Integrated Development Environment (IDE), ancillary tools, and the operating system. We validate DAT through observational studies, surveys, visualizations, and statistics.
Out of more than 20 studies completed with DAT, we highlight three, demonstrating DAT's impact across Meta:
\begin{enumerate}
    \item A controlled experiment found that the introduction of mock types led to a 14\% DAT improvement, marking a first in linking language-level features to productivity gains.
    \item A post-hoc case study showed quantification of productivity gains from a new feature in the React Compiler, auto-memoization, resulting in 33\% DAT reduction.
    \item An estimation based counterfactual model showed estimated efficiency gains from code sharing as a $> 50$\% improvement across frameworks, saving thousands of engineering hours annually.
\end{enumerate}

Overall, the contributions of this paper are three-fold:
\begin{enumerate}
    \item We propose a novel, yet straight-forward way to measure development velocity using a combination of readily available telemetry sources including version control and the IDE.
    \item We describe how the use of this metric has enabled rigorous experimentation on development productivity at industry scale, yielding new research findings such as that types do indeed make developers move faster.
    \item We argue for a more scientific mindset to internal product development, going beyond qualitative evidence.
\end{enumerate}

\section{Related Work \& Background}

\begin{figure*}[tb]
\centering
\includegraphics[width=\textwidth]{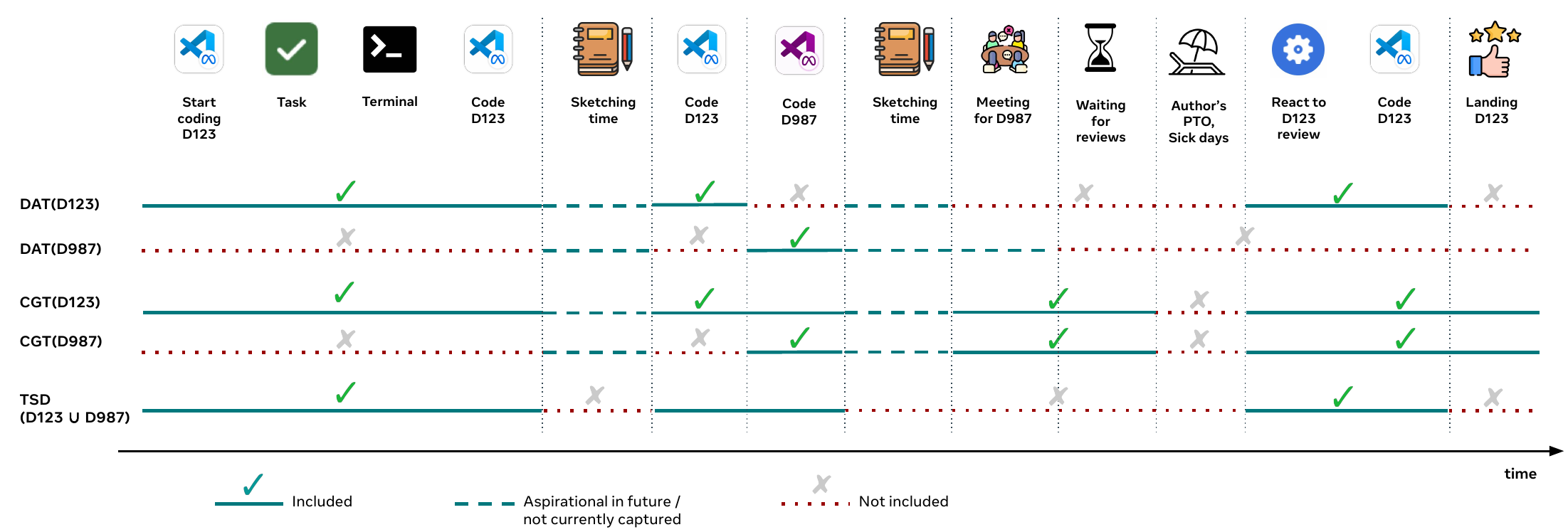}
\caption{Example illustration of the most relevant DAT-covered and uncovered activities for two hypothetical diffs D123 and D987, and a comparison with other related metrics Code Gestation Time (CGT) and Time Spent by Diff (TSD).}
\label{fig:dat_overview}
\end{figure*}

\subsection{Continuous Experimentation}
``Continuous Experimentation'' (CE) refers to the practice of regularly running experiments to validate one's hypotheses. While this approach has become widespread in large, software-intense organizations and has defined much of the engineering work in Big Tech~\cite{kohavi2009online}, it is not ubiquitous~\cite{mantyla2022viability}. Challenges to adoption include a lack of resources and knowledge on CE, particularly in startups.

Melegati et al. establish a ``CE maturity model'' with five stages of adoption and
implementation~\cite{melegati2020xpro}. When our work commenced, the engineering culture regarding CE in Meta's infrastructure organization was somewhat similar to a startup: while Meta as a whole was naturally aware of CE (stage 1), it required a cultural mind shift to embrace it for Infra (stage 2). An organic need driven by lower observed output combined with leadership mandates initiated this process. To instill the new mindset and work processes throughout the organization, a particularly rigorous validation was necessary (stage 3), convincing key stakeholders. We derived results by focusing on a generally accepted measure (DAT), and integrating it into Meta's already validated experimentation infrastructure (stage 4). Dedicated data scientists performed much of this work, leading to fast productionization of these results, including publishing the achieved wins in this paper (stage 5+).


\subsection{Developer Productivity Metrics}
\label{sec:related_metrics}

Productivity metrics in Software Engineering have an over 50 year long history~\cite{brooks1974mythical}, with dozens of different metrics proposed. Prominent product metrics include as lines of code~\cite{jones1976quantitative} or change count~\cite{meli2001measuring}. Traditionally, one core problem with many of these metrics is their lack of meaning on ``actual productivity'' or failing to take into account the complexity of some tasks (a one-liner bug fix can be more complex than creating a green-field system and may have a higher impact). As a solution, function point analysis~\cite{symons1988function,abran2002function} has been suggested, which delegates judgment of the complexity to a third-party, supported by guidelines. This is costly, may still be inaccurate, and has come out of fashion, perhaps due to the short iteration cycles in modern development that often obviate planning. Instead, a stronger emphasis on the developer experience (``DevEx'') has emerged in industry~\cite{fagerholm2012developer,noda2023devex}. At the same time, the scientific community has started to focus on (self-)perceived productivity~\cite{meyer2017characterizing,cheng2022improves}. 

Recent works such as SPACE~\cite{space} and DX Core4~\cite{core4} provided an umbrella framework on how to think about software development productivity at large, but defer the exact metric choices to their users. The chosen metrics then often focus on easily measurable aspects of the development process, rather than the complexity of the authoring. For example, output metrics such as lines of code or change count emphasize the result of development but fail to capture the effort and complexity involved in the authoring lifecycle. Review time metrics, while good for measuring collaboration and delayed reviews, are only a sub-part of the authoring process. Another concern arises from the combination of many metrics into a construct that is now hard to reason about and act on.

Existing productivity metrics are thus either gameable, subjective, impossible to obtain accurately, a poor proxy of productivity, too complex, or some combination of the above~\cite{kitchenham1997counterpoint,petersen2011measuring,symons1988function,huijgens2014replicated}. One scholar even suggest abandoning productivity measurement altogether~\cite{ko2019we}. The underlying challenge is that productivity measurement for knowledge workers, of which software engineers are a prime example~\cite{ford2017characterizing}, is a fundamentally unsolved problem~\cite{ramirez2004measuring}. 

In this paper, we adopt a pragmatic approach, focusing on accurate time measurement for authoring. ``Time spent coding'' best aligns with developers' own productivity perceptions~\cite{beller2020mind}, is objective, and can be measured at scale. To the best of our knowledge, no such metric for authoring has been formalized yet, since it would need to encompass all the different tools, workflows, and idiosyncrasies involved in authoring code. DAT does exactly this by measuring the active time taken to develop a diff while enabling insight into the components that impact it, such as IDE features, libraries, and programming languages. For instance, feature experimentation using DAT enables teams to assess the impact of specific improvements on developer velocity.

We propose DAT as a conceptually simple, objective metric that is not merely theoretical, but of high practical value. At Meta, DAT supersedes Time Spent by Diff (TSD) and Code Gestation Time (CGT), illustrated in \Cref{fig:dat_overview}. The figure shows engineering-related activities for two diffs (\emph{D123} and \emph{D987}) across different tools, and in subsequent rows, which activities each metric covers. The illustrated sequential development workflow starts with coding on \emph{D123} and ends with landing it, with various activities (among them, coding on the unrelated \emph{D987}) interspersed. DAT accurately covers both diffs' core engineering, specifically in IDEs, while it cannot (yet) capture off-screen sketching time. TSD averaged coding time in a given period by the number of diffs published, but cannot differentiate between diffs \emph{D123} and \emph{D987} in the same period, as it lacks this level of precision. Thus, TSD was unable to run diff-level experiments. CGT, a composite metric, measured time from coding start to diff landing, mixing active engineering with technical time to release, double-counting overlapping engineering time (for \emph{D987}), complicating interpretation and lacked validation.

\subsection{Development Flow at Meta}

\subsubsection{A diff: Meta's pull request equivalent}
A \emph{diff}, e.g., \emph{D123}, represents a single code change including a natural text title, summary, and test plan~\cite{beller2023learning}, similar to a pull request~\cite{gousios2014exploratory}. Almost all code-based changes end up as (ideally) self-contained diffs that can stack on top of each other. A typical way to make diffs is by creating commits in an IDE such as VS Code~\cite{vscode}, using Sapling, Meta's integrated, platform-independent version control client for multiple version control systems~\cite{goode2014scaling,sapling}. However, there are a number of other modalities for making diffs, including bespoke tools~\cite{murali2024ai}.
Code review of all diffs happens on a custom version of the open-source Phabricator tool~\cite{kudrjavets2022mining,beller2023learning}.

\subsubsection{Diff-grain productivity measurements}
Many productivity experiments make sense on the diff level, meaning that one developer can appear both in the A and B group for different diffs, naturally removing one of the biggest confounders. For example, a diff-level experiment could be to test the DAT effects of a new Continuous Integration (CI) signal, which, as one among many hundreds of CI signals, would not influence the developers' core workflow or even be noticed if absent in the future. Similarly, switching out the base language model from LLaMA-33B to LLaMA-65B~\cite{touvron2023llama} would likely also be somewhat opaque operation to a developer.
Other experiments, though, need to be on the cohort level---it makes little sense and is practically impossible to switch one developer back-and forth between two VS Code versions with completely different user interfaces, for example. A diff-based measurement is still essential in these cases, as the diff level provides normalization for many other confounders such as length, complexity, language, framework, or testing.  Overall, these characteristics make diffs a natural and high-coverage measurement unit for productivity and experimentation purposes.

\section{Diff Authoring Time (DAT)}

\subsection{Architecture}
Diff Authoring Time (DAT) aims to capture all active human developer time spent on a particular diff, split by the different roles involved (e.g., author and reviewer). As \Cref{fig:dat_overview} shows, the key component of DAT is accurate representation of time spent in the IDE. DAT builds upon this to cover all diff-related activities and tools.  Naturally, time off the computer, such as sketching, whiteboarding, or meeting time, will always be difficult to track. 

While there is an ongoing scientific debate on humans' ability to multitask~\cite{koch2018cognitive,skaugset2016can}, active human work on a computer tends to be highly sequential, even if such individual sequences can be extremely short (e.g., rapid application switching often happens in $<1s$): there can only be one application in focus at a given time, and such activity is atomically related to one diff only.

This gives DAT advantageous properties that lend themselves to validation: it is non-overlapping (per user), meaning that two diffs by the same author must not have any overlapping time. As a corollary, DAT cannot exceed 24 hours per day (though it is, in practice, much shorter), can sensibly be aggregated (as opposed to CGT), and the sum of DAT in one day is bounded by TSD. Thus, DAT is a high-confidence, unobjectionable floor value of development time spent that is attributable to a diff.  In fact, the comparison with TSD can reveal (1) improvement potential for DAT's matching algorithm in cases where time is part of the larger TSD, but should also be part of DAT, or (2) potential for overall efficiency improvement, as coding-related activities in TSD correctly uncaptured in DAT might not have resulted in a productive output artifact (a diff), thus potentially representing ``wasteful'' effort. 
The goal for DAT is to be used purely as a productivity metric for tooling and workflows, not for evaluating people. Access and privacy restrictions ensure it can never be used in performance reviews, for example.

\subsection{Algorithm}
DAT consists of a main algorithm, precise matches, and a number of heuristics to capture additional time.

\subsubsection{Core DAT algorithm: Precise matches}

\begin{figure}[tb]
\centering
\includegraphics[width=0.5\textwidth]{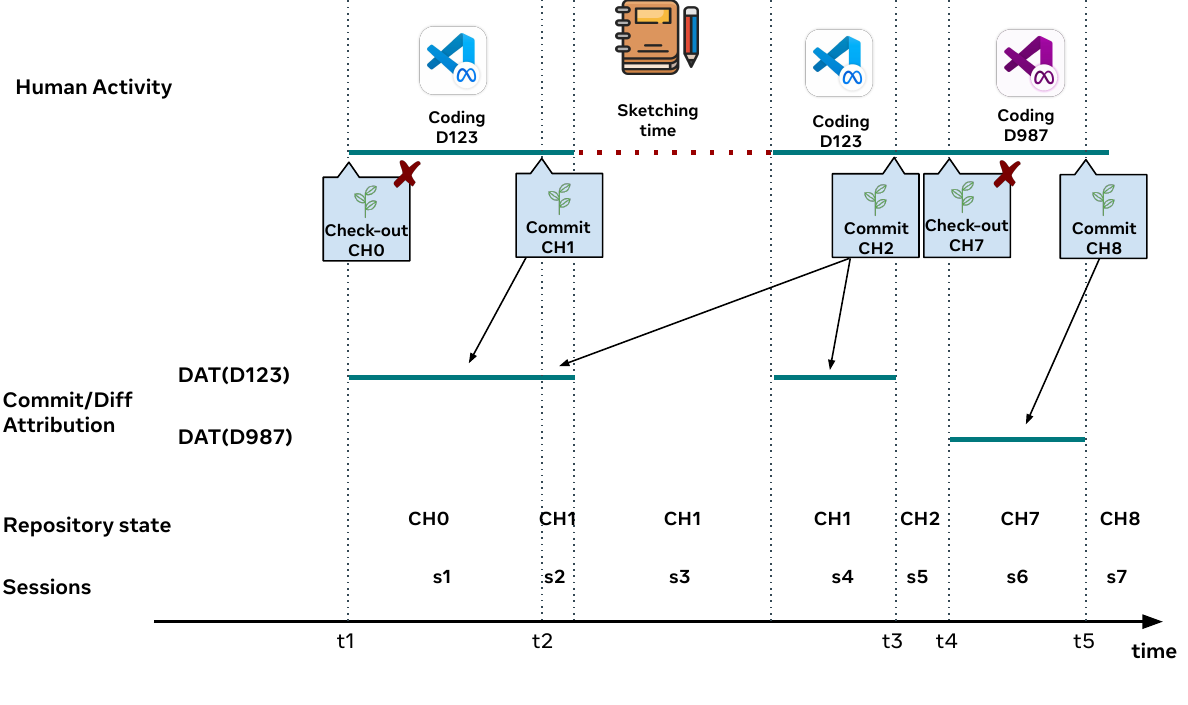}
\vspace{-0.8cm}
\caption{DAT's core commit matching algorithm.}
\label{fig:dat_matching}
\end{figure}

As the first two rows in \Cref{fig:dat_overview} depict, the core authoring of a diff starts with the initial trace of an activity that we can connect with this diff. Typically, this is the start of the first IDE session---most often in VS Code---in which the first commit is made that is then later connected to that diff. DAT includes only the time the engineer spends actively working on this diff, as the core DAT algorithm priorities precise, high-confidence matching---it omits IDE activity that does not result in a commit, or activity for commits that do not result in a diff. It only operates on tools that allow such high confidence attribution: IDEs and the Phabricator Diff review tool.

We obtain basic activity and foreground application (the application window in focus) through OS-level telemetry. We then correlate it to commit events using telemetry from the Sapling plugin inside the IDE. For authors of a diff, we only consider \emph{modifying} commit events (e.g., {\tt commit} or {\tt amend}); for reviewers, we only consider \emph{non-modifying} events (e.g., a checkout). Meta's version control generates a new, unique commit hash even when modifying an existing commit and it does not allow rewriting of history. The Sapling plugin in the IDE talks to a host-wide Sapling server to which all Sapling client calls are routed. This has the advantage of tracking the underlying repository as a singleton, meaning that any interaction with it, be it from inside or outside the IDE (e.g., also from the command line), is captured. 

\Cref{fig:dat_matching} depicts the fundamental principle of DAT's precise diff-to-session matching algorithm using the example of two diffs D123 and D987 that are developed sequentially in the same VS Code session: on top, the figure shows human activity as work sessions, along with point-in-time interactions with version control. It also shows the current head commit (i.e., the state of the repository from {\tt sl status}, similar to {\tt git status}). The figure depicts the principle that any commit $CH_x$ is attributed on the time axis to the IDE session $s_{x-1}$ that precedes it. Thus, at its core, the DAT algorithm performs a 1-left-shift of the underlying commits with regard to the telemetry sessions. For example, the commit $CH1$ made at time $t2$ is attributed to the VS Code session $s1$ that happened from $t1$ to $t2.$ This is because any change developed starting from $t1$ until $t2$ will become part of the commit $CH1$ made at time $t2$. Similarly, both $s2$ and $s4$ are mapped to $CH2$. $s2$ is interrupted by the offline activity $s3$ (``sketching''), which does not count toward DAT as OS-level telemetry indicates inactivity (e.g., no mouse movements or key input due to a coffee break). Timeout parameters have been chosen in line with existing work~\cite{beller2017developer}.

There are two corner cases: 
\begin{enumerate}
    \item Any time a developer opens VS Code, an initial automatic checkout happens ($CH0$ in \Cref{fig:dat_matching}). As the developer does not modify this pre-existing commit, the automatic commit checkout event filtered away by the DAT algorithm. Similarly, DAT removes all manual, non-modifying checkouts (e.g., the later checkout of the pre-existing commit $CH7$) via explicit filtering. The commit $CH8$, which becomes $CH7$'s successor, is attributed to D987. 
    \item For the ``trailing session'' $s7$, there is no commit mapping because any work done there would not actually be saved as a commit.
\end{enumerate}

We now have a precise commit-to-session mapping (hence the name ``precise matches''), but not yet a mapping to diffs. This is trivial to achieve, as version control provides us with a list of all commits that make up a diff.

\subsubsection{Additional DAT Heuristic: Anchor-DAT}

DAT offers a heuristic beyond precise matches. It was developed to easily capture more of the time surrounding diff creation, especially in tools that have no notion of diffs or commits, such as interactive playgrounds---e.g., a SQL interface. The idea is to map \emph{coding activity} preceding a precise match to the same diff as the precise match, recognizing that related activities mostly co-occur sequentially.

For instance, in \Cref{fig:dat_overview}, work in the external terminal application cannot be precisely matched due to lack of Sapling integration, but additional heuristics map it to D123 because it precedes a precise match VS Code session to \emph{D123}.
This conservative method captures anchor sessions leading up to a diff, which are otherwise hard to trace. Activity is captured only for a limited time before a precise match (a few minutes) and only if it involves a list of manually vetted coding-related tools. This list is shared with TSD, making DAT a subset of the duration covered by TSD. Whenever we refer to DAT in the following, we refer to the combination of precise matches and Anchor-DAT.



\subsection{Validation}
\label{sec:validation}

\subsubsection{User Experience Research Study}
\label{sec:uxr}
Our first step in validating DAT was to create a ground truth (GT) dataset of how much time developers spend on a diff. This allowed us to assess DAT's accuracy by comparing it with GT data and observing trends as we improved the algorithms and tweaked its main parameters (namely, the list of to-be-supported internal coding tools, the anchor length, and inactivity timeout).

With a team of User Experience (UX) researchers we conducted a study in which we screen-recorded  developers working on a diff from inception to landing, then transcribed their actions and the duration of them. Finally, we compared the results to the extracted DAT sessions on a by-second basis. To counteract biases from screen-recording and volunteering, we used a stratified random selection of developers and diffs for broad coverage, ensuring that different strata were present in our sample (e.g., small diffs, stacked diffs, different languages and IDEs, different developer roles).

We analyzed $n = \sim 20$ diffs, achieving an average accuracy of over 90\%, meaning the computed DAT was on average within $\pm 5\%$ of its GT value. In later iterations of the algorithm, DAT's accuracy surpassed human labeling for many of the analyzed diffs, as developers rapidly switched between tools, activities, and diffs. These sometimes sub-second context switches made it hard for the UX annotators to keep track of developers' work items at all times, especially when multiple diffs were developed in parallel. When we revisited the source video for the diverging GT transcriptions, we often found DAT to be correct.

\subsubsection{Large Scale Survey}

\begin{figure}[tb]
\centering
\includegraphics[width=\linewidth]{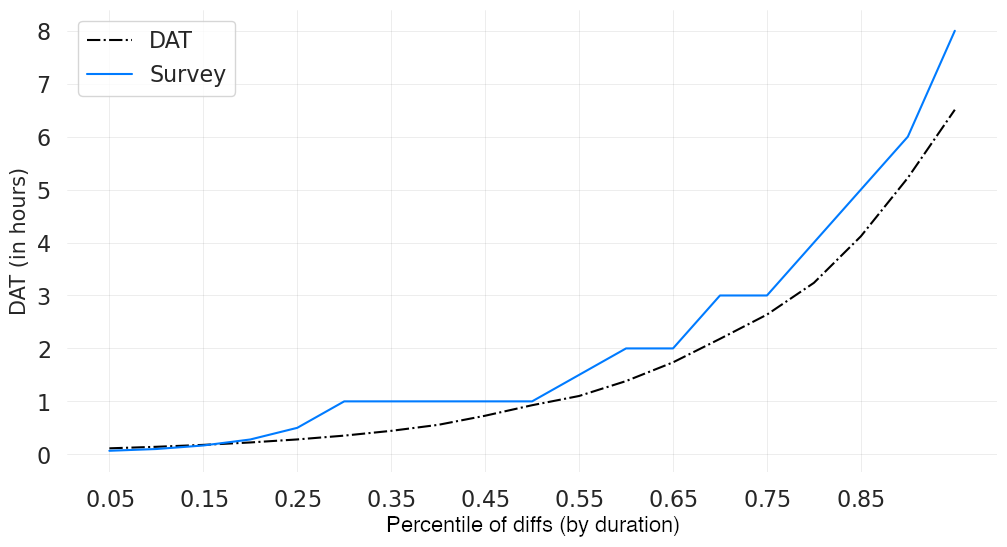}
\caption{Survey-estimated (solid) vs. actual DAT values show the small systematic lower bound of DAT compared to human estimates.}
\label{fig:dat_comparison}
\end{figure}

\begin{figure}[tb]
    \centering
    \includegraphics[width=\linewidth]{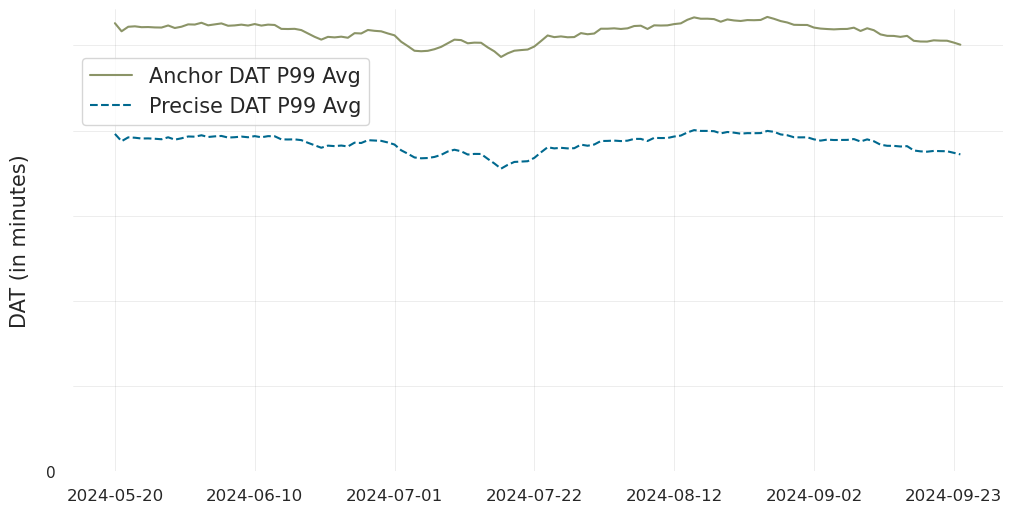}
    \caption{99th percentile winsorized average DAT for different algorithms, across a range of several months. The figure clearly shows the stability of DAT as a top-line metric.}
    \label{fig:trendline}
    \end{figure}

We found that acceptance of an important top-line metric depends on how well it matches experience and intuition. Therefore, we conducted a large-scale survey asking developers to estimate the time spent on a diff. This one-question survey was integrated into the Phabricator tool, appearing once a diff was shipped to ensure responders had the full development context fresh in their minds.

We received responses for 968 unique diffs to compare with DAT values. As \Cref{fig:dat_comparison} shows, DAT and survey data generally align, with developer-reported times being slightly higher, a trend also seen in the UX study (\Cref{sec:uxr}). We observe that DAT remains a floor for all measured DAT percentiles.

\subsubsection{Descriptive Statistics}
We performed two main validations for DAT at an aggregate level---diff coverage and diff duration. These validations were intended to verify that DAT covers the expected number of diffs (diff coverage) and that the overall time spent in these diffs yields reasonable values. As a sanity check, the coverage and duration validations were done to surpass pre-defined goals. The duration validations also included comparison between the summed-up DAT for a given time period and user and their respective TSD (cf. \Cref{sec:related_metrics}), where $\sum DAT \leq TSD$

We performed several tests based on our expectation of how much Anchor-DAT should increase the precise matches. First, we expected anchor sessions' durations to be higher than precise match durations. Second, we expected that the sum of all precise matches and anchor session durations never be larger than the sum of all TSD for a user during a particular time window.  Any exceptions would only be possible if there is a mistake in the data or our algorithms, for example due to double-counting time. A final demand on a top-line metric is to be stable over time---\Cref{fig:trendline} shows this.

Before releasing DAT, we ensured it passed all these checks. We found that DAT covers >95\% of all eligible diffs. DAT's duration has a right-skewed distribution, which means a small number of extreme values can significantly impact the mean. To account for longer diffs while still being sensitive to changes in the distribution, we use the 99th percentile winsorized mean to report on Anchor-DAT, which yielded plausible results consistent with the other validations.

\subsubsection{Time-Line Visualization}

\begin{figure*}[htbp]
\centering
\includegraphics[width=1\textwidth]{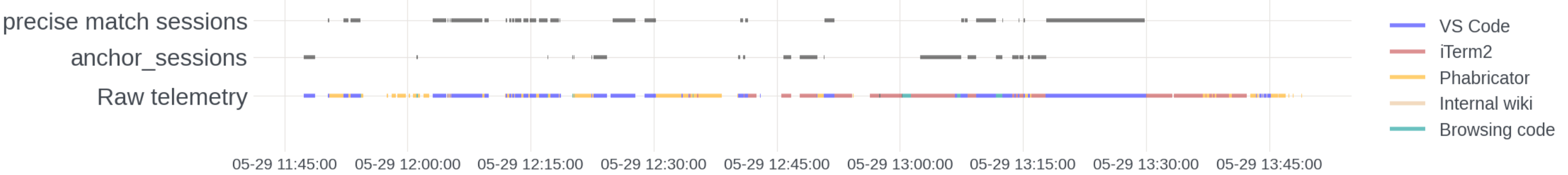}
\caption{A sample timeline of activities (raw telemetry, bottom row), and different DAT algorithms.}
\label{fig:dat_timeline}
\end{figure*}

\Cref{fig:dat_timeline} visualizes raw telemetry (bottom row) from a two hour time frame, how it is mapped to an example diff's precise DAT (top row), and how auxiliary tools are covered via Anchor-DAT. DAT uses raw telemetry sessions, including start and stop times of coding tools, shown in the bottom row. DAT tries to map each session to a diff. The raw telemetry sessions already include noise-removal, such as pauses for inactivity (e.g., around 12:45).
During Anchor-DAT development, we also cross-validated with authors to ensure extrapolations were sensible. These checks helped refine our understanding, re-evaluate heuristic assumptions, and adjust parameters like time thresholds.

\section{Experiment: Typed Mocking}
\label{sec:typed_mock}



\begin{figure*}[htbp]
\centering
\begin{subfigure}[t]{\linewidth}
\begin{lstlisting}[basicstyle=\ttfamily\small,escapechar=|,numbers=left]
 // Existing test case
 class AgeTest extends MetaTest {
   public function testYoungCantVote(): void {
     self::mock(Database::queryUserData<>, ($_user) ==> dict["name" => "Alice", "age" => 16]);
     expect(Application::canUserVote("Alice"))->toEqual(false);
   }
 }

+class UserData {
+   function __construct (public string $name, public int $age ) {}
+}

-class Database { function queryUserData(string $user): dict<string, mixed>; }
+class Database { function queryUserData(string $user): UserData; }

 class Application {
   function canUserVote(string $user): bool {
     $user_data = Database::queryUserData($user);|\label{line:query}|
-    return (int)$user_data['age'] >= 18;
+    return $user_data->age >= 18;
   }
 }
\end{lstlisting}

\caption{An exemplary code change (``diff'') introducing a new type, {\tt UserData}.}

\vspace{1cm}
\label{lst:preamble}
\end{subfigure}
\begin{subfigure}[t]{0.48\linewidth}
\begin{lstlisting}[basicstyle=\ttfamily\small,escapechar=|]
// Testing framework, old, untyped mock function
class MetaTest {
  function mock(mixed $fn, mixed $callable);
}
\end{lstlisting}
\caption[b]{The old mocking function leads to a {\tt TypehintViolation} in {\tt AgeTest} at runtime.}
\label{lst:untyped_mock_error}
\end{subfigure}
\hspace{0.02\linewidth} 
\begin{subfigure}[t]{0.48\linewidth}
\begin{lstlisting}[basicstyle=\ttfamily\small,escapechar=|]
// Testing framework, new, typed, mock function
class MetaTest {
  function mock<T>(FunctionRef<T> $fn,
  T $callable): void;
}
\end{lstlisting}
\caption{The new mocking function leads to a static type (lint) error in {\tt AgeTest} at development time.}
\label{lst:typed_mock}
\end{subfigure}
\caption[b]{Comparison of untyped and typed mocking APIs.}
\label{fig:mocking_comparison}
\end{figure*}

\subsection{Context}
Hack is one of Meta's server-side languages. Originating from PHP, it combines a JIT-based runtime~\cite{guilherme2018hhvm} with a gradual, object-oriented type system, providing
static correctness guarantees similar to TypeScript for JavaScript. Hack developers can depend on commonly shared frameworks, including mocking for testing~\cite{spadini2017mock}. While CI forces test execution for every diff, developers can additionally run them locally if they so wish.

Hack code is statically typed to varying degrees. In the presence of suitable type annotations, Hack has the ability to detect certain type errors through static analysis, which are reported to developers directly in VS Code. In addition to static analysis, the Hack runtime enforces the arguments passed to functions, and ensures that they match the declared parameter types, throwing a dynamic {\tt TypehintViolation}-exception in case of a mismatch.
In particular, the standard mocking framework used to expose an untyped API, using the catch-all {\tt mixed} type. As a result, its use did not enjoy any static guarantees. Values flowing through {\tt mixed} could encounter runtime enforcement boundaries, potentially leading to runtime crashes.

\subsection{Problem Statement}
Listing~\ref{lst:preamble} presents an example diff, with unaltered lines for context and lines prefixed with {\tt +} or {\tt -} added or removed, respectively. In this diff, a developer improves the {\tt Application} class (lines 16ff.) to be more object-oriented, introducing a new {\tt UserData} class (lines 9--11), and altering existing code (lines 13 and 19) to make efficient use of it (line 14). However, the developer might not be aware of the already existing test (lines 1--7) for it, for example, because the test code might live in a completely different folder structure and is not easily discoverable.

In the case of the old, untyped mocking framework (Listing~\ref{lst:untyped_mock_error}), they would learn that they had broken a test only after CI ran,  introducing significant delays: since the mocked method body now returns a {\tt dict} value where {\tt UserData} is expected, the Hack runtime throws a {\tt TypehintViolation}. This leads to an additional cycle of checking out a diff, reproducing and debugging the unexpected failure, coming up with the right fix, and verifying it. The lack of static types for mocking has proven to be a hindrance to productivity, as indicated by the disproportionately high number of {\tt TypehintViolation}-exceptions from CI logs.

To avoid this degraded experience, we built a typed variant of the existing mocking framework. Listing~\ref{lst:typed_mock} shows the new  typed {\tt mock} function. The Hack type checker will immediately report a type mismatch as a type error in the developer’s IDE because the closure type is incompatible, without the need for the developer to know about or execute any potentially existing tests.



\subsection{Methodology}
Our experiments showed that we would be able to automatically migrate the majority (82\%) of calls to the untyped mocking API onto the new typed variant ({\it migratable call sites}). The remainder of call sites (or {\it unmigratable call sites}) exercised the mocking API in unconventional ways and could not be auto-migrated.

We first migrated a random selection of half of the migratable call sites ($\sim$169,000 or 42\% of all call sites in the codebase). During three weeks in 2024, we categorised $\sim$10,000 diffs:

\begin{itemize} 
\item \emph{Control Group (53.1\% of diffs)}: Diffs that modified {\it at least one unmigrated, but no partially or fully migrated files}. 
    \item \emph{Test Group (24.5\% of diffs)}: Diffs that modified {\it at least one fully migrated, but no partially migrated or unmigrated files}. 
    \item \emph{Mixed Group (22.4\% of diffs)}: Diffs that modified {\it at least one migrated and one unmigrated file, or at least one partially migrated file}. 
\end{itemize}

To estimate the effect of typed mocking, we employ a two-tailed Welch's {\it t}-test, testing on the mean DAT difference between control and test groups. We discarded the mixed group as it contained diffs that touched both migrated and unmigrated files, making it ambiguous how to treat them. Because bigger diffs naturally require more coding time, we controlled for the diff size confounder using stratification.

\subsection{Results}

\begin{table}[tb]
\centering
\caption{Typed Mocking Experiment results for {\it t}-tests with diff size as stratum.}
\begin{tabular}{cccr}
\toprule
\# of files changed & \% DAT saved & $\frac{|Control|}{|Test|}$ & $p-value$ \\
\midrule
1        & $\sim22\%$ & $2.5$  & $<0.001$  \\ 
2        & $\sim22\%$ & $2.3$ & $<0.001$ \\
3        & $\sim19\%$ & $2.2$  & $0.003$  \\
4        & $\sim24\%$ & $2.0$  & $0.001$  \\
\midrule
Variable & $\sim14\%$ & $2.2$  & $<0.001$  \\
\bottomrule
\end{tabular}
\label{tab:typed_mocking_stratified_tests}
\end{table}

\Cref{tab:typed_mocking_stratified_tests} summarizes the results of the stratified {\it t}-tests, revealing that use of the typed mocking API significantly reduced the time to write diffs, even after controlling for the number of files in each diff. The table shows the percentage of DAT saved, the relative control and test group sizes, and the p-value for each diff size category; diff size is defined as the number of files changed in a diff.

Diffs of all size categories showed statistically significant DAT reductions of approximately $14\%$  DAT ($p < 0.001$, $n = \sim 10,000$). This was particularly prevalent for diff sizes up to four files, where diffs that used the typed mocking API showed an approximate 19--24\% reduction in DAT over the control.
We surmise that the inclusion of larger diffs (with more than four files changed), which might only tangentially touch mocking-related files, dilutes the larger savings seen in smaller diff size strata.

After seeing these positive results, Meta decided to fully migrate all 82\% migratable call sites to typed mocking. 
While the results of this experiment have helped to quantify the direct effect of a Hack language feature and promoted a decision to roll it out across the code base, the wider decision-making processes in the organization also benefited.
To the best of our knowledge, this is the first time a controlled, large-scale experiment demonstrated real-world productivity benefits of types in industry.  Moreover, this result demonstrates that investing in low-latency static analysis tooling to shift correctness signals early in the development cycle can lead to substantial improvements in developer productivity, especially in large code bases.

\section{Case Study: React's Auto-Memoization}
\label{sec:react}

\subsection{Context}
React is Meta's JavaScript framework for developing dynamic user interfaces~\cite{gackenheimer2015introduction}. Since its open-source release in 2013, it has become a standard for web and mobile apps~\cite{react}. React components generate user interfaces based on the application's current state. When the state changes, React updates the interface by  re-evaluating components. This can be computationally intense. To optimize performance, developers often use memoization. In a sense, memoization is the front-end equivalent of caching. It prevents re-computing and re-drawing of unchanged components. This is especially important since components often form a deeply nested hierarchy and updating an unmemoized, but also unchanged parent component could cause a cascade of unnecessary re-renders.

\subsection{Problem Statement}
To avoid this, React developers can make their React components manually support memoization. This involves writing code to save component states manually; it has to be done for every single component. Manual memoization, similar to caching, not only causes a lot of non-trivial overhead  unrelated to the actual feature engineers are implementing, but also introduces bugs and performance regressions, when done incorrectly~\cite{SavonaZhang2023IdiomaticReact}. For example, just to signal to the compiler that a certain component {\tt ExampleView} supports memoization, a function's signature had to look like this:
\begin{lstlisting}[basicstyle=\ttfamily\tiny,escapechar=|,numbers=none]
export default (React.memo<Props>(ExampleView): React.ComponentType<Props>); 
\end{lstlisting}
instead of a simple 
\begin{lstlisting}[basicstyle=\ttfamily\tiny,escapechar=|,numbers=none]
export default ExampleView;
\end{lstlisting}

The actual implementation logic inside the components was often dozens or hundreds of lines long.

For this reason, the React team set out to develop the next iteration of its bespoke React client-side compiler, React Forget. Its main feature is auto-memoization, which abstracts away manual memoization for the developer. With it, the compiler can determine the correct level and depth of memoization and export it automatically, eliminating the need for long export signatures and  complex memoization logic.

\subsection{Methodology and Results}
\begin{figure}[tb]
    \centering
    \includegraphics[width=1\linewidth]{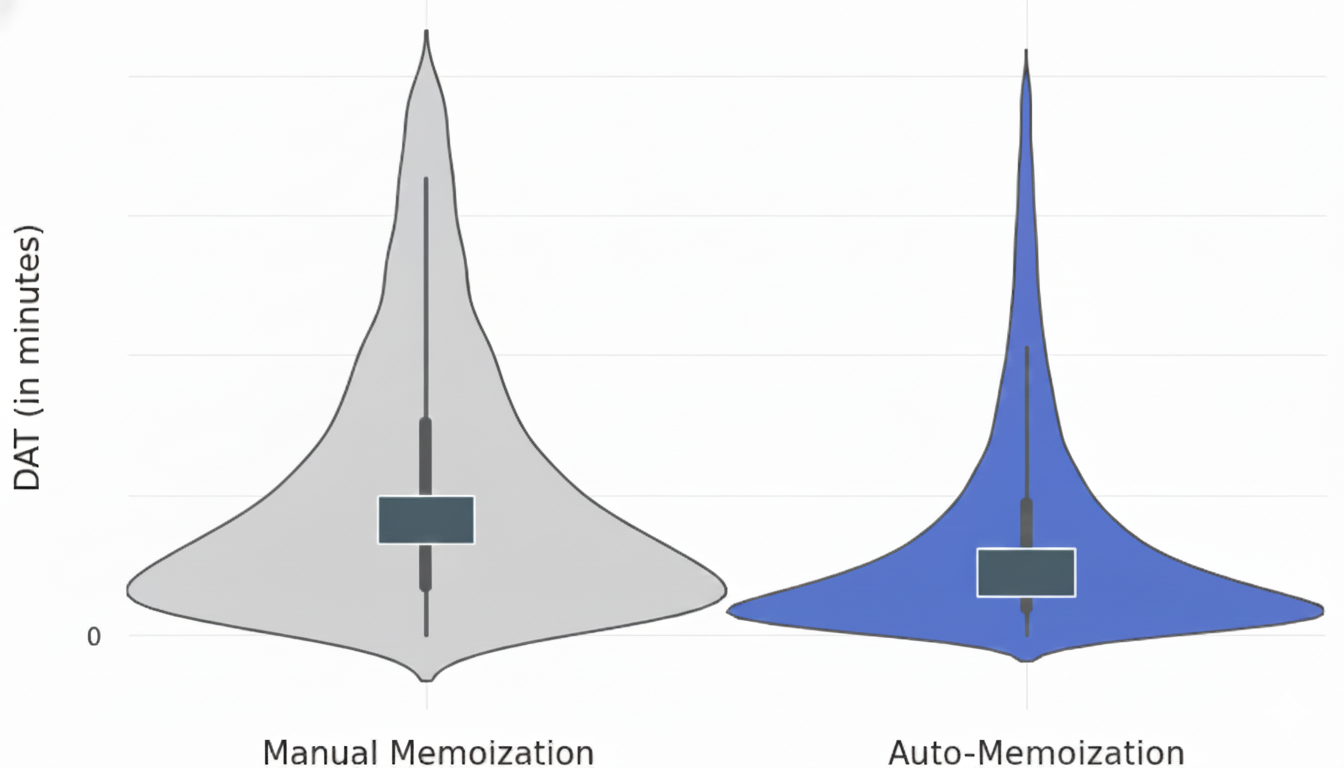}
    \caption{Beanplots of DAT for diffs with manual (left) and automatic memoization (right) in a  subset of React diffs.}
    \label{fig:Beanplot}
\end{figure}

While hard evidence regarding the usefulness of React Forget existed on operational metrics such as app performance and bug rates, it was initially impossible to evaluate its other purported aim, namely the impact on development productivity. With DAT, we could scientifically address it in a case study.

To obtain an initial, intuitive understanding of the difference the new React compiler might make to development velocity, we first compared diffs with auto-memoization to those without. Results ($n >> 10,000$ diffs) in \Cref{fig:Beanplot} show that the DAT average, median, quartiles, and skew are significantly higher (i.e., worse) for diffs with manual memoization.

Based on these encouraging results, to obtain a precise difference between groups, and to take into account confounders, we built a mixed-effects regression model. This included lines of code, developer tenure, organization, time (modeled as month of the year), and others as its independent variables. It models \%$\Delta \mbox{DAT}$ as its output, to measure differences in mean DAT duration. 
As a post-hoc case study with non-randomized data, we use the Wasserstein distance to measure the difference between groups. It calculates the minimal transport distance between two probability distributions~\cite{wasserstein} and is robust to noise and outliers, making it ideal for comparing distributions with different shapes or scales. 

The final result of the mixed-effects model put \%$\Delta \mbox{DAT}$ between diffs with manual and auto-memoization at 33\%. These DAT savings from the introduction of the React Forget compiler turned out greater than anticipated by domain experts.

\section{Case Study: Quantifying Efficiency Gains from Code Sharing}

\subsection{Context}
Meta's development frameworks enable cross-platform code sharing, allowing code to be written once and deployed across multiple apps and platforms, reducing redundant efforts. One well-known example of such a framework is---React. Using code-sharing frameworks may save time compared to writing platform-specific code, because code only needs to be written once, but can be deployed to a number of different platforms. Developers report feeling more efficient with these frameworks, yet quantifying these gains was impossible before DAT.

\subsection{Methodology}
To quantify $DAT_{saved}$, we subtract actual DAT development time from an estimated counterfactual DAT without code sharing.

Counterfactual DAT aims to provide a realistic estimation of the development time that would have been required if code sharing had not been employed. While actual DAT for code-shared diffs is directly measurable, the counterfactual development time cannot be observed since these diffs were never created. Instead, we estimate it by identifying the platforms (such as Android) and apps (such as Instagram or Threads) targeted by each code-shared diff. For each platform, we compute a trimmed mean DAT from unshared diffs and normalize it by diff size, categorizing diffs into terciles based on lines of code modified. 

The total counterfactual DAT is then the sum of the mean DAT for each targeted platform-app combination in the code-shared diff, controlled for diff size. Consequently, $DAT_{saved}$ can be negative if a code-shared diff incurs overhead, potentially due to the complexity of using or getting used to the shared libraries. Conversely, a higher count of targeted platform-app pairs increases the potential for actual DAT savings.

To illustrate this with a simple, hypothetical example, consider a code-shared diff developed using a framework that supports cross-platform deployment. This diff, with a total DAT of 1 hour, is deployed across four apps: Instagram Android, Instagram iOS, Facebook Android, and Facebook iOS.  In contrast, implementing the same change independently for each app would require four separate diffs (this is the counterfactual to code sharing). Of course, we cannot simply multiply the actual DAT by this number to arrive at the counterfactual, but need to approximate it much closer, taking into account the idiosyncratic productivity in each different ecosystem. Querying the population diffs of comparable size might yield an average DAT of 0.9 hours for iOS and 0.6 hours for Android diffs (these are not the actual values). Therefore, the total counterfactual DAT is 3 hours (2 × 0.9 + 2 × 0.6), resulting in an estimated $DAT_{saved}$ of 2 hours. 

\subsection{Results}
We initially applied this method to two major code-sharing frameworks, covering thousands of code-sharing diffs and counterfactual diffs monthly (total $n = \sim 50,000$). Results indicate that for these two frameworks alone, we save thousands of hours of DAT annually, a relative improvement to non-shared development of over 50\%. We then expanded the methodology with more parameters to other frameworks. Results generally pointed to large engineering savings from code sharing on a continuous basis.

\section{Threats to Validity}
\subsection{Construct Validity}
To mitigate implementation errors, we performed a quadruple validation of DAT (see \Cref{sec:validation}), including large-scale surveys and end-to-end tests with GT. Moreover, we plan to increase accuracy by supporting more tools and IDEs with precise matches. For instance, we have already seen the precision improvements of the addition of Android Studio as a natively supported IDE.

\subsection{Internal Validity}
As a top-line metric, DAT strikes a balance between support for all engineering workflows at the company (recall), and correctness for each covered diff (precision). These objectives are often at odds, as the example of Anchor-DAT shows: extrapolation increases tool coverage at the expense of potentially covering activity not related to this diff.

To cater to these different demands, we designed a highly accurate base algorithm for DAT (precise matches), and augment it with a heuristic. DAT consumers can choose their individual precision-recall trade off based on their requirements.
DAT integrates with Meta's internal experimentation framework Deltoid, which automatically handles power analysis, effect size calculation, experiment tracking, multiple simultaneous experiment adjustments, and reporting~\cite{fb:scale2014}, ensuring reported effects uphold high rigor. 

\subsection{External Validity}
We believe DAT can apply to a multitude of stakeholders outside of Big Tech. It might be easier to obtain buy-in in other organizations than Meta, which needs to support a plethora of products, environments, and developer preferences. For example, an organization that develops a Windows app would only need to support the exact tools used in their specific development flow.

While experimentation is most effective with enough data points in the control and test group, it is often easier to run case studies, which can also provide convincing evidence. Power analysis can reveal that sometimes a few dozen developers can be sufficient to detect effect sizes of the magnitude we present in this paper~\cite{fitzner2010sample}. It is also not strictly necessary to have full control over the complete engineering workflow, as one is typically interested in a specific aspect of it with an experiment. There are many organizations that have the means to acquire such data and drive improvements based on it, such as consultancies, tool vendors, and open source projects (for example, the Eclipse IDE for Java surpassed 12 million monthly downloads in 2023~\cite{eclipse}). Even some researchers~\cite{beller2017developer} attract developer bases large enough to run sufficiently powered experiments. Finally, even in the absence of many users, can the mental model of experimentation-guided software development help shape how developers build products.

\section{Discussion}

\subsection{Studies at Industry Scale}
\subsubsection{Experiments}
Randomized controlled trials and controlled experiments are often called the gold standard in science~\cite{manzi2012uncontrolled}. Unfortunately, they continue to be a rarity in Software Engineering~\cite{sjoberg2005survey,stol2018abc}.

While a relatively large body of such studies exists in the space of types~\cite{hanenberg2010experiment,endrikat2014api,kleinschmager2012static}, understandably, one limitation of many previous studies was the reliance on students as experiment subjects~\cite{carver2004issues} and the inability to do like-for-like comparisons between control and experiment groups. This might be a reason why studies reached contradictory conclusions about the effectiveness of types. Our experiment here suffers no such limitations and is, to the best of our knowledge, the first one to be conducted in a large-scale, state-of-the-art, industrial setting with real developers on real-world problems. Partially contradicting previous findings~\cite{hanenberg2010experiment}, our results show strong productivity improvements when using static types. It is possible that the nature of our experiment, namely the introduction of typed mocking in a large-scale system, is more comparable to previous studies which favored types as system or task complexity increased~\cite{stuchlik2011static,endrikat2014api,gao2017type}. While this hypothesis seems plausible, we need more rigorous experiments in different settings to assess whether static types hinder ad-hoc, quick development, and are actually a productivity improvement in larger projects. It is also unclear where the threshold between these two extremes lies.

\subsubsection{Case Studies}
For code sharing and compiler improvements, there exists a vast body of largely anecdotal or post-hoc empirical analyses touting their benefits: These range from cost reduction to increased developer satisfaction~\cite{haefliger2008code,bhattacharya2011assessing,mockus2007large}. These works corroborate our findings in spirit, but seldom contain concrete efficiency gains, stem mostly from Open-Source Systems, lack a focus on (time-based) developer productivity, or are limited in size. The studies in this paper address these gaps by establishing concrete productivity gains, working on industrial systems, and containing a large amount of data.

\subsection{Practical Implications}
Practitioners in platform and framework teams often ask how they can affect developer productiveness and satisfaction. The studies in this paper show that these teams can have significant impact on the productivity of the engineers in their organization, assuming they focus their efforts on the right projects. By contrast, other studies performed with DAT and not reported in this paper showed that small, incremental UX improvements often did not significantly affect DAT. 

This distinguishes internal developer experimentation from external consumer experimentation, where changing the position of a button by a few pixels can yield measurable changes in user behavior~\cite{airbnb}. The fidelity of DAT might not lend itself to such micro optimizations.

This could be attributed to the fact that developers are highly cognitively engaged in their work and constantly striving to find optimizations themselves. Their scope and creativity are broader than the often confined ecosystems in the consumer-facing product world, which means that capturing workflows in an accurate metric is much harder. An example would be developers that continue to use their favorite editor, even if there is only one ``officially'' supported editor, and paste their edits in. The DAT for such copy-and-pasted diffs would be artificially low, and would introduce noise.

\subsection{Dealing with Small Effect Sizes}
Another challenge is establishing statistical significance in DAT, especially when working with small sample sizes. To overcome this, experiments often need to run for extended periods to account for variability in workflows and usage patterns. 

In such cases, engineering teams might show impact by improving an operational metric more directly affected by their changes, for example, number of successful searches. While a statistically significant, positive change in this metric might not yield statistically significant effects on DAT, the team could rely on a previously demonstrated, principal correlation between that metric and DAT to claim productivity improvements.

\subsection{Embracing an Experiment-first Culture}
Setting up an experiment requires time and rigor, and may require a shift in the behavior of engineers, who might not be used to this mindset. As an example, the experiment in \Cref{sec:typed_mock} involved a staged migration of {\tt mock} call sites, to ensure DAT would be correctly measured. This additional effort for the sake of an experiment cannot always be justified. To design code in a DAT-first way can significantly alter the feature itself.

As a benefit this mindset aligns the experiment-driven consumer-facing product development workflow with how internal Infrastructure teams operate, thus making it easier to collaborate or switch between teams. Thus, DAT can facilitate greater inter-team mobility and more uniform workflows in the company.

\subsection{Going Beyond Velocity}
This paper demonstrates the broad applicability of DAT in practice,  enabling Software Engineering-related case studies and experiments, from language features and compiler improvements to a generalized method to estimate framework efficiency gains.  While DAT offers insights into velocity, there are other important aspects to software development such as quality~\cite{cavano1978framework} or developer satisfaction~\cite{storey2019towards}. Our experiments and case studies have shown, though, that these often go together: a DAT improvement frequently co-occurs with other desired properties, such as more static safety guarantees through typed mocking in~\Cref{sec:typed_mock} or higher developer satisfaction.

Our studies have established that language and framework improvements (e.g., the React compiler improvements in~\Cref{sec:react}) provide large velocity improvements.  However, for an end-to-end analysis, their initial development efforts need to be balanced against the later DAT savings, ensuring the benefits outweigh the initial added development overhead. Fortunately the initial development cost can be measured with DAT, too, providing a net benefit analysis of these features. Having performed over 20 DAT experiments and case studies, we are now in a position to inform planning and prioritization of development in a rigorous, data-driven way.

\subsection{Avoiding Metric-Blindness}
DAT has provided key insights into developer velocity. By precisely capturing the previously opaque authoring time, DAT enabled a paradigm shift for Infrastructure developers, bringing them into a more data-driven and objective world.

Yet, it is important to recognize that metric movement should not become a condition or barrier to pursuing revolutionary ideas. Some improvements may not result in (measurable) DAT reductions, either because they cannot be measured or because no valid experiment can be set up. It remains paramount that we as software builders continue to listen to our expert opinion on future innovations as well as current workflow hindrances, next to assessing more large-scale sentiment and qualitative feedback from fellow developers. We must maintain a balance between leveraging DAT for data-driven decisions and fostering a culture where innovative ideas can thrive. Consider the example of coding agents that ingest a development task and work on it in a human-like way~\cite{hassan2024rethinking}. Due to their current speed limitations, they might initially have adverse impact on DAT (i.e., a developer can perform the same actions in a much shorter DAT-equivalent than the agent), though their long-term trajectory looks extremely promising.

\section{Conclusion \& Future Work}
We conceptualized DAT as a practical productivity metric, implemented it, and used it successfully in more than 20 projects across Meta, of which we highlight three in this paper:

\begin{enumerate}
    \item  Mock types in Hack, denoting a 14\% DAT improvement; to the best of our knowledge, this was the first, large-scale, real-world experiment accounting for relevant confounders that demonstrated productivity improvements by static typing;
\item Auto-memoization in the React compiler, showing a 33\% DAT improvement;
\item Increased use of code sharing, showing a $> 50\%$ DAT improvement, saving thousands of hours annually.
\end{enumerate}

In conclusion, DAT has enabled Meta's Infrastructure organization to adopt a metric-first development approach, aligning it closer with product teams. This way, DAT instigated a paradigm shift in how Infrastructure engineers work, report, and plan. DAT also paved the way toward rigorous experimentation on long-standing software engineering debates, such as ``do types make development more efficient.''

\label{sec:immediate}
We are continuously evolving DAT by increasing coverage horizontally and vertically.
Horizontally, we will expand DAT beyond diffs to include other artifacts such as documents and tasks, creating a general time-based measurement framework for experimentation.
Vertically, we aim to cover more tools and IDEs natively, having to rely less on heuristics.

As AI advances in software development~\cite{khemka2024toward}, we believe DAT will remain relevant. Coding
agents learn from human interactions, and DAT could not only be used to measure continuous improvements, but also guide the optimization (or loss) function
toward higher productivity. Finally, the shift to a scientific, experiment-driven internal engineering culture will hopefully have long-lasting impact on engineers and organizations.



\bibliographystyle{ACM-Reference-Format}
\bibliography{main}

\end{document}